\documentclass[12pt]{article}
\usepackage[cp1251]{inputenc} %- для WiN
\usepackage[english, russian]{babel} %- для рус-англ. переноса
\usepackage{amsmath, latexsym,  amssymb, bm,  graphics,
amsfonts, array, eucal}%,newlfont, bm}% linedraw}
\usepackage[dvips]{graphicx}
\makeatletter

\renewcommand{\Re}{\mathop{\rm Re\,}}
\renewcommand{\Im}{\mathop{\rm Im\,}}

\makeatother

\newcommand{\be}{\begin{center}}
\newcommand{\ee}{\end{center}}

\begin{document}
\renewcommand{\abstractname}{\ Abstract}
\renewcommand{\refname}{\begin{center} REFERENCES\end{center}}
\newcommand{\mc}[1]{\mathcal{#1}}
\newcommand{\E}{\mc{E}}
\thispagestyle{empty}
\large

\begin{center}
\bf Generation of longitudinal electric current
by the transversal electromagnetic field in classical and quantum
degenerate plasma
\end{center}

\begin{center}
  \bf A. V. Latyshev\footnote{$avlatyshev@mail.ru$} and
  A. A. Yushkanov\footnote{$yushkanov@inbox.ru$}
\end{center}\medskip

\begin{center}
{\it Faculty of Physics and Mathematics,\\ Moscow State Regional
University, 105005,\\ Moscow, Radio str., 10A}
\end{center}\medskip

\begin{abstract}
The analysis of nonlinear interaction of transversal
electromagnetic field with degenerate collisionless
classical and quantum plasmas is carried out.
Formulas for calculation electric current in degenerate collisionless
classical and quantum plasmas are deduced.
It has appeared, that the nonlinearity account leads to occurrence
of  longitudinal electric current directed along a wave vector.
This second current is orthogonal to the known transversal  current,
received at the classical linear analysis.
Graphic comparison of density of  electric current for classical
degenerate Fermi plasmas and Fermi---Dirac plas\-mas (plasmas with any
degree of degeneration of electronic gas) is carried out.
Graphic comparison of density of electric current for
classical and quantum degenerate plasmas is carried out.
Also comparison of dependence of density of electric
current of quantum degenerate plasmas
from dimen\-sion\-less wave number at various values
of dimensionless frequency of oscillations of
electro\-mag\-netic field is carried out.


electron gas
\end{abstract}

\begin{center}
\bf  Introduction
\end{center}

In the present work formulas for calculation electric current
in classical and quantum collisionless degenerate plasmas are deduced.

At the solution of the kinetic Vlasov equation describing
behaviour of classical degenerate plasmas, we consider as in decomposition
distribution functions, and in decomposition of quantity
of the self-conjugate electro\-mag\-ne\-tic field the quantities proportional
to square of intensity of an external electric field.

At the solution of the kinetic equation with Wigner integral, describing
behaviour of quantum degenerate plasmas, we consider as in decomposition
distribution functions, and in decomposition of Wigner integral
the quantities proportional to  square
of potential of an external electromagnetic field.

At such nonlinear approximation it appears, that an electric current
has two nonzero components. One component of electric
current is directed along intensity of electric field (in
classical plasma) or along potential of  electro\-mag\-ne\-tic field
(in quantum plasma). This  component of  electric field
precisely the same, as well as in the linear analysis. It is "transversal" \,
current.
Thus, in linear approximation we receive the known
expression of the transversal electric current.

The second nonzero of electric current component has the second
order of infinitesimality concerning with quantity intensity of
electric field (in case of classical plasma) or the second
order of infinitesimality of quantity concerning with potential quantity of
electromagnetic field (in case of quantum plasma).
The second component of electric current  is directed along the wave
vector both in classical and in quantum plasma. This current
is perpendicular to the first  component. It is "longitudinal" \,  current.

So, vector expression of an electric current consists from two composed,
orthogonal each to other.
The first composed, linear on intensity of electric field, is
known classical expression of an electric current along the electric
fields. The second composed represents an electric current,
which is propor\-tio\-nal to an intensity square
of electric field. The second current
is perpendicular to the first and it is directed along the wave
vector. Occurrence of the second current comes to light the spent
the nonlinear analysis of interaction of  electromagnetic field
with classical and quantum plasma.

Nonlinear effects in plasma are studied already long time
\cite{Gins} -- \cite{Shukla2}.

In works \cite{Gins} and \cite{Zyt} nonlinear effects are studied in
plasma. In work \cite{Zyt} the nonlinear current was used, in
particulars, in probability of questions of disintegration
processes. We will note,
that in work \cite{Zyt2} it is underlined existence
nonlinear current along a wave vector
(see the formula (2.9) from \cite{Zyt2}).

Quantum plasma was studied in works \cite{Lat1} -- \cite{Lat8}.
Collisional quantum plasma has started to be studied in work
of Mermin \cite{Mermin}. Then quantum collisional plasma
it was studied in our works \cite{Lat2} -- \cite{Lat6}.

The collisional quantum plasma with variable frequency of
collisions in our works \cite{Lat4} and \cite{Lat5} was studied.
In works \cite{Lat7} and \cite{Lat8} was investigated
generation of  longitudinal current by the transversal
electric field in classical and quantum Fermi---Dirac plasma
 \cite {Lat7} and in Maxwellian plasma
\cite{Lat8}.

\begin{center}
  \bf 1. {The case of classical plasma}
\end{center}

Let us show, that in case of the classical plasma
described by the Vlasov equation, longitudinal current is generated
and we will calculate its density. On
existence of this current was specified more half a century ago
\cite{Zyt2}.
We take  Vlasov equation describing behaviour
of collisionless plasmas
$$
\dfrac{\partial f}{\partial t}+\mathbf{v}\dfrac{\partial f}{\partial
\mathbf{r}}+
e\bigg(\mathbf{E}+
\dfrac{1}{c}[\mathbf{v},\mathbf{H}]\bigg)
\dfrac{\partial f}{\partial\mathbf{p}}=0.
\eqno{(1.1)}
$$

Electric and magnetic fields are connected with the vector
potential by equalities
$$
\mathbf{E}=-\dfrac{1}{c}\dfrac{\partial \mathbf{A}}{\partial
t},\;\qquad \mathbf{H}={\rm rot} \mathbf{A}.
$$

Let us consider, that the intensity of electric and magnetic fields vary
harmoniously
$$
{\bf E}={\bf E}_0e^{{\bf i(kr-\omega t)}},\qquad
{\bf H}={\bf H}_0e^{{\bf i(kr-\omega t)}}.
$$

The wave vector we will direct along an axis $x $: $ {\bf k} =k (1,0,0) $, and
intensity of electric field we will direct along an axis $y $:
$ {\bf E} =E_y (0,1,0) $.

Hence,
$$
\mathbf{E}=-\dfrac{1}{c}\dfrac{\partial \mathbf{A}}{\partial
t}=\dfrac{i\omega}{c}\mathbf{A},\qquad A_y=-\dfrac{ic}{\omega}E_y,
$$
$$
{\bf H}=\dfrac{ck}{\omega}E_y\cdot(0,0,1),\qquad
{\bf [v,H}]=\dfrac{ck}{\omega}E_y\cdot (v_y,-v_x,0),
$$
$$
e\bigg(\mathbf{E}+\dfrac{1}{c}[\mathbf{v},\mathbf{H}]\bigg)
\dfrac{\partial f}{\partial\mathbf{p}}=
\dfrac{e}{\omega}E_y\Big[kv_y\dfrac{\partial f}{\partial p_x}+
(\omega-kv_x)\dfrac{\partial f}{\partial p_y}\Big].
$$

Let us operate with the method of consecutive approximations, considering
as small parameter the quantity of intensity of electric field.
Let us rewrite the equation (1.1) in the form
$$
\dfrac{\partial f^{(k)}}{\partial t}+v_x\dfrac{\partial f^{(k)}}{\partial x}=
$$
$$
=-
\dfrac{eE_y}{\omega}\Bigg[kv_y\dfrac{\partial f^{(k-1)}}{\partial p_x}+
(\omega-kv_x)\dfrac{\partial f^{(k-1)}}{\partial p_y}\Bigg],\qquad k=1,2.
\eqno{(1.2)}
$$ \bigskip

Here $f^{(0)}$ is the absolute Fermi distribution,
$$
f^{(0)}=f_0=\Theta(\E_0-\E),\qquad \E=\dfrac{mv^2}{2},\qquad
\E_0=\dfrac{mv_0^2}{2},
$$
$\Theta(x)$ is the unit step of Heaviside,
$\E$ is the electron energy, $\E_0$ is the electron energy on Fermi
surface,  $p_0=mv_0$ is the electron momentum on Fermi surface,
$v_0$ is the electron velocity on Fermi surface.

We notice that
$$
[\mathbf{v,H}]\dfrac{\partial f_0}{\partial \mathbf{p}}=0,
$$
because
$$
\dfrac{\partial f_0}{\partial \mathbf{p}}\sim \mathbf{v}.
$$

We search for the solution as  first approximation in the form
$$
f^{(1)}=f_0+f_1,
$$
where
$$
f_1\sim E_y,\qquad E_y\sim e^{i(kx-\omega t)}.
$$

In this approximation the equation (1.2) becomes simpler
$$
\dfrac{\partial f_1}{\partial t}+v_x\dfrac{\partial f_1}{\partial x}=
-eE_y\dfrac{\partial f_0}{\partial p_y}.
$$

From here we receive
$$
-i(\omega-kv_x)f_1=-\dfrac{eE_y}{m}\cdot \dfrac{\partial f_0}{\partial v_y}.
\eqno{(1.3)}
$$

From (1.3) we obtain
$$
f_1=-\dfrac{eE_y}{m}\cdot\dfrac{\partial f_0/\partial v_y}{\omega-kv_x}.
\eqno{(1.4)}
$$

Here
$$
\dfrac{\partial f_0}{\partial v_y}=-\delta(\E_0-\E)mv_y.
$$

In the second approximation for the solution of the equation (1.2)
we search in the form
$$
f^{(2)}=f^{(1)}+f_2=f_0+f_1+f_2,
$$
where
$$
f_2\sim E_y^2,\qquad E_y^2\sim e^{2i(kx-\omega t)}.
$$

Let us substitute $f^{(2)} $ in the equation (1.2).
Considering the equation (1.3), we come to the equation
$$
\dfrac{\partial f_2}{\partial t}+v_x\dfrac{\partial f_2}{\partial x}=
-\dfrac{eE_y}{\omega m}\Bigg[kv_y\dfrac{\partial f_1}{\partial v_x}+
(\omega-kv_x)\dfrac{\partial f_1}{\partial v_y}\Bigg].
$$ \bigskip

From this equation we obtain
$$
f_2=-\dfrac{ieE_y}{2m\omega(\omega-kv_x)}
\Bigg[kv_y\dfrac{\partial f_1}{\partial v_x}+
(\omega-kv_x)\dfrac{\partial f_1}{\partial v_y}\Bigg]=
$$
$$
=-\dfrac{e^2E_y^2}{2m^2\omega(\omega-kv_x)}\Big[kv_y
\dfrac{\partial}{\partial v_x}\Big(\dfrac{\partial f_0/\partial v_y}
{\omega-kv_x}\Big)+\dfrac{\partial^2 f_0}{\partial v_y^2}\Big].
\eqno{(1.5)}
$$

The distriburion function in second approximation is constructed
$$
f=f^{(2)}=f^{(0)}+f_1+f_2,
\eqno{(1.6)}
$$
where $f_1, f_2$ are given by equalities (1.4) and (1.5).

Let us find electric current density
$$
\mathbf{j}=e\int \mathbf{v}f \dfrac{2d^3p}{(2\pi\hbar)^3}=
e\int \mathbf{v}(f_1+f_2) \dfrac{2d^3p}{(2\pi\hbar)^3}.
\eqno{(1.7)}
$$
From equalities (1.4) -- (1.6) it is visible, that the vector
of  current density has two nonzero components
$$
\mathbf{j}=(j_x,j_y,0).
$$

Here $j_y$ is the density of transversal current,
$$
j_y=e\int v_yf \dfrac{2d^3p}{(2\pi\hbar)^3}=
e\int v_yf_1 \dfrac{2d^3p}{(2\pi\hbar)^3}.
\eqno{(1.8)}
$$

This current is directed along an electric field, its density
is defined only by the first approximation of distribution function.
The second approximation   of distribution function the
contribution to current density does not bring.

The density of transversal current is defined by equality
$$
j_y=\dfrac{2e m^3}{(2\pi\hbar)^3}\int v_y f_1 d^3v.
$$

For density of longitudinal current according to its definition it is had
$$
j_x=e\int v_xf\dfrac{2d^3p}{(2\pi\hbar)^3}=
e\int v_xf_2\dfrac{2d^3p}{(2\pi\hbar)^3}=
\dfrac{2e m^3}{(2\pi\hbar)^3}\int v_xf_2d^3v.
$$

By means of (1.6) and (1.5) from here it is received, that
$$
j_x=-\dfrac{e^3E_y^2 m}{(2\pi\hbar)^3\omega}\int
\Bigg[kv_y
\dfrac{\partial}{\partial v_x}\Big(\dfrac{\partial f_0/\partial v_y}
{\omega-kv_x}\Big)+\dfrac{\partial^2 f_0}{\partial v_y^2}\Bigg]
\dfrac{v_xd^3v}{\omega-kv_x}.
\eqno{(1.9)}
$$

In the second integral from (1.9) internal integral on $P_y $ is equal
to zero
$$
\int\limits_{-\infty}^{\infty}\dfrac{\partial^2 f_0}
{\partial v_y^2}dv_y=0.
$$

In the first integral from (1.9) internal integral on $P_x $
is calculated in parts
$$
\int\limits_{-\infty}^{\infty}\dfrac{\partial}{\partial v_x}
\Big(\dfrac{\partial f_0/\partial v_y}{\omega-kv_x}\Big)
\dfrac{v_xdv_x}{\omega-kv_x}=
-\omega \int\limits_{-\infty}^{\infty}
\dfrac{(\partial f_0/\partial v_y)dv_x}{(\omega-kv_x)^3}.
$$

Hence, equality (1.9) becomes simpler
$$
j_x=\dfrac{e^3E_y^2mk}{(2\pi\hbar)^3}\int
\dfrac{(\partial f_0/\partial v_y)v_yd^3v}{(\omega-kv_x)^3}.
\eqno{(1.10)}
$$

We notice that
$$
\int\limits_{-\infty}^{\infty}v_y\dfrac{\partial f_0}{\partial v_y}dv_y=
v_0\Theta(v_0-v)\Bigg|_{v_y=-\infty}^{v_y=+\infty}-
\int\limits_{-\infty}^{\infty}\Theta(v_0-v)dv_y.
$$

Equality (1.10) is reduced now to integral
$$
j_x=-\dfrac{e^3E_y^2mk}{(2\pi\hbar)^3}\int
\dfrac{\Theta(v_0-v)d^3v}{(\omega-kv_x)^3}.
\eqno{(1.11)}
$$

The three-dimensional integral is equal
$$
\int \dfrac{\Theta(v_0-v)d^3v}{(\omega-kv_x)^3}=
\int\limits_{v^2<v_0^2}\dfrac{d^3v}{(\omega-kv_x)^3}=
$$
$$
=\int\limits_{-v_0}^{v_0}\dfrac{dv_x}{(\omega-kv_x)^3}
\iint\limits_{v_y^2+v_z^2<v_0^2-v_x^2}dv_ydv_z=
\pi\int\limits_{-v_0}^{v_0}\dfrac{(v_0-v_x^2)dv_x}{(\omega-kv_x)^3}=
$$
$$
=\dfrac{\pi}{k_0^3}\int\limits_{-1}^{1}\dfrac{(1-\tau^2)d\tau}
{(\Omega-q\tau)^3}.
$$

Hence, the longitudinal current is equal
$$
j_x=-\dfrac{e^3E_y^2mk\pi}{(2\pi\hbar)^3k_0^3}
\int\limits_{-1}^{1}\dfrac{(1-\tau^2)d\tau}
{(\Omega-q\tau)^3}.
\eqno{(1.12)}
$$

Here $q $ is the dimensionless wave number, $ \Omega $ is the
dimensionless frequency of oscillations of electromagnetic field,
$$
q=\dfrac{k}{k_0},\qquad \Omega=\dfrac{\omega}{k_0v_0}.
$$

Let us find numerical density (concentration) of particles of the plasma,
corresponding to Fermi---Dirac distribution
$$
N=\int \Theta(v_0-v)\dfrac{2d^3p}{(2\pi\hbar)^3}=
\dfrac{2 m^3}{(2\pi\hbar)^3}\int \Theta(v_0-v)d^3v=
\dfrac{k_0^3}{3\pi^2},
$$
where $k_0$ is the  Fermi wave number, $k_0=\dfrac{mv_0}{\hbar}$.

In expression before integral from (1.12) we will allocate the plasma
(Lang\-muir) frequency
$$
\omega_p=\sqrt{\dfrac{4\pi e^2N}{m}}
$$
and number density (concentration) $N$,
and last we will express through Fermi wave number. We will receive
$$
{j_x}^{\rm long}=-{E_y^2}\dfrac{3e\Omega_p^2}{k_0p_0}\dfrac{k}{32\pi}
\int\limits_{-1}^{1}\dfrac{(1-\tau^2)d\tau}{(\Omega-q\tau)^3},
$$
where
$\Omega_p=\dfrac{\omega_p}{k_0v_0}=\dfrac{\hbar\omega_p}{mv_0^2}$
is the dimensionless plasma frequency.

The previous equality we will copy in the form
$$
j_x^{\rm long}=J_{\rm c}(\Omega,q)\sigma_{\rm l,tr}kE_y^2,
\eqno{(1.13)}
$$
where $\sigma_{\rm l,tr}$ is the longitudinal--transversal conductivity,
$J_{\rm c}(\Omega,q)$ is the dimen\-sion\-less  part of current,
$$
\sigma_{\rm l,tr}=
\dfrac{e\hbar}{p_0^2}\Big(\dfrac{\hbar \omega_p}{mv_0^2}\Big)^2=
\dfrac{e}{p_0k_0}\Omega_p^2,
$$
$$
J_{\rm c}(\Omega,q)=-\dfrac{3}{32\pi}
\int\limits_{-1}^{1}\dfrac{(1-\tau^2)d\tau}{(\Omega-q\tau)^3}.
$$

The integral from dimensionless part of  current is calculated according to
known Landau rule that is equivalent to equality
$$
J_{\rm c}(\Omega,q)=-\dfrac{3}{32\pi}\lim\limits_{\varepsilon\to +0}
\int\limits_{-1}^{1}
\dfrac{(1-\tau^2)d\tau}{(\Omega-i\varepsilon-q\tau)^3}.
\eqno{(1.14)}
$$

If we introduce the transversal field
$$
\mathbf{E}_{\rm tr}=\mathbf{E}-\dfrac{\mathbf{k(Ek)}}{k^2}=
\mathbf{E}-\dfrac{\mathbf{q(Eq)}}{q^2},
$$
then equality (1.13) can be written down in invariant form
$$
\mathbf{j}^{\rm long}=
J_{\rm c}(\Omega,q)\sigma_{\rm l,tr}{\bf k}{\bf E}_{tr}^2
=J_c(\Omega,q)\sigma_{\rm l,tr}\dfrac{\omega}{c}[{\bf E,H}].
$$

Let us pass to consideration of the case of small values of wave
number. From expression (1.12) at small values of wave number it is received
$$
j_x^{\rm classic}=-\dfrac{4e^3E_y^2mk}{24\pi^2\hbar^3k_0^3\Omega^3}=
-\dfrac{e\omega_p^2E_y^2k}{8\pi(k_0p_0)^2 k_0p_0\Omega^3}=
%$$
%$$=
-\dfrac{1}{8\pi \Omega^2}\cdot\sigma_{\rm l,tr} kE_y^2.
\eqno{(1.15)}
$$

Let us notice, that in the case of degenerate  plasmas and real
and imaginary parts of
longitudinal current beyond all bounds increase at $q\to \Omega $.
This singularity indicate the sharp vanishing of
distribution function of degenerate plasmas at once behind Fermi's surface.

At regimes of non-degenerate  Fermi---Dirac plasmas, close to
degenerate regime (the quantity $ \alpha $ is large, but it is finite),
i.e. at the temperatures close to zero, but not
equal to zero, singularity in the point $q =\Omega $ is absent.

From Figs. 1 and 2 it is visible, that at  increase
of dimensionless chemical potential $ \alpha $ velocity of
growth of the real part of the longitudinal current increases also.

In the limit at $ \alpha\to + \infty\; (T\to +0) $
the quantities $ \Re J_c (\Omega, q) $ and $ \Re J_c (\Omega, q) $ becomes
explosive in the point $q =\Omega $.
It also is the case of degenerate plasmas.

For graphics of the real and imaginary parts of the longitudinal
current in Fermi---Dirac  plasma on Figs. 1 -- 4 the formula
of longitudinal current from our work \cite{Lat7} is used.
In this formula transition to dimensionless quantities
by electron velocity $v_0$ on Fermi's surface is spent.
This formula looks like
$$
j_x^{\rm quant}=J_c(\Omega,q)\sigma_{\rm l,tr}kE_y^2,
$$
where
$$
\sigma_{\rm l,tr}=\dfrac{e\Omega_p^2}{p_0v_0},\qquad
\Omega_p=\dfrac{\omega_p}{k_0v_0},
$$
$$
l_0(\alpha)=\int\limits_{0}^{\infty}\ln(1+e^{\alpha-\tau^2})d\tau,
$$
$$
J_c(\Omega,q)=\dfrac{1}{16\pi l_0(\alpha)}\int\limits_{-\infty}^{\infty}
\dfrac{\ln(1+e^{\alpha(1-\tau^2)})d\tau}{(q\tau -\Omega)^3}.
$$

Imaginary part of the dimensionless current in Fermi---Dirac plasma
is calculated according to the previous
equalities and Landau rules under the formula
$$
\Im J_c(\Omega,q)=-\dfrac{1}{32l_0(\alpha)q^3}[\ln(1+e^{\alpha(1-\tau^2)})]''
\Bigg|_{\tau=\Omega/q}=
$$
$$
=\dfrac{\alpha}{16l_0(\alpha)q^3}\cdot
\dfrac{1+(1-2\alpha\tau^2)e^{-\alpha(1-\tau^2)}}
{[1+e^{-\alpha(1-\tau^2)}]^2}\Bigg|_{\tau=\Omega/q}.
$$

With imaginary part of the dimensionless current in
degenerate plasma things are more difficult. According to Lahdau
rule imaginary part is calculated under the formula
$$
\Im J_c(\Omega,q)= \dfrac{3}{32q^3}\Big\{\begin{array}{c}
0, q<\Omega, \\ 1, q>\Omega \end{array}\Big\}.
$$

On the other hand, if to integrate twice in parts
integral from (1.14), we receive
$$
\int\limits_{-1}^{1}\dfrac{(1-\tau^2)d\tau}{(q\tau-\Omega+i\varepsilon)^3}=
\dfrac{2(\Omega-i\varepsilon)}{q^2[(q^2-(\Omega-i\varepsilon)^2]}-\dfrac{1}{q^2}
\left|\dfrac{q-\Omega}{q+\Omega}\right|-
$$
$$
-\dfrac{i}{q^3}
\Big[\arctg\Big(\dfrac{q-\Omega}{\varepsilon}\Big)+
\arctg\Big(\dfrac{q+\Omega}{\varepsilon}\Big)\Big].
$$

Let us consider the quantity
$$
A=\dfrac{2(\Omega-i\varepsilon)}{q^2[q^2-(\Omega-i\varepsilon)^2]}\sim
\dfrac{2(\Omega-i\varepsilon)}{q^2[(q^2-\Omega^2+2i\Omega\varepsilon]}.
$$

If the parameter $q $ runs to $ \Omega $, we receive
$$
\Im A=-\dfrac{1}{\Omega^2 \varepsilon}.
$$

If now the parameter $ \varepsilon $ runs to zero, we find, that
$$
\Im A\to \infty.
$$

Let us consider that $q \neq \Omega $, and then if the parameter
$ \varepsilon \to 0$ it is received, that
$$
\Im A \to 0.
$$

It means, that limits in (1.14) on $q\to 0$ and $ \varepsilon\to 0$
are non-commutativity.

As it was already specified, it mean
that the sharp vanishing of distribution function of degenerate
plasma at once behind Fermi's surface.
To eliminate this lack of the description of degenerate plasma
is possible or
"diffusion"\, Fermi--surface, or consideration such regimes,
close to degenerate regime (at $ \alpha\to +\infty $), or introduction
in consideration of electron collisions and cal\-cu\-la\-tion so
named "weakly collision"\,  limit at $q\to 0$.

On Figs. 3 and 4 are presented graphics in which it is found out
dependence of real  (Fig. 3) and imaginary (Fig. 4) parts
longitudinal current from quantity of dimensionless wave number $q $ at
various values of quantity of dimensionless frequency of oscillations
of electromagnetic field $ \Omega $.

It appears, that at
increase $ \Omega $ amplitude excursion of values of the real
and imaginary parts of longitudinal current sharply decreases.
A minimum of imaginary part $q_{min} $ is near to the point
$q =\Omega $. With increase
of quantity of chemical potential $ \alpha $ this minimum moves in
point $q =\Omega $: $q_{min} \to \Omega $.

%\clearpage

\begin{figure}[h]\flushleft
\includegraphics[width=18.0cm, height=14cm]{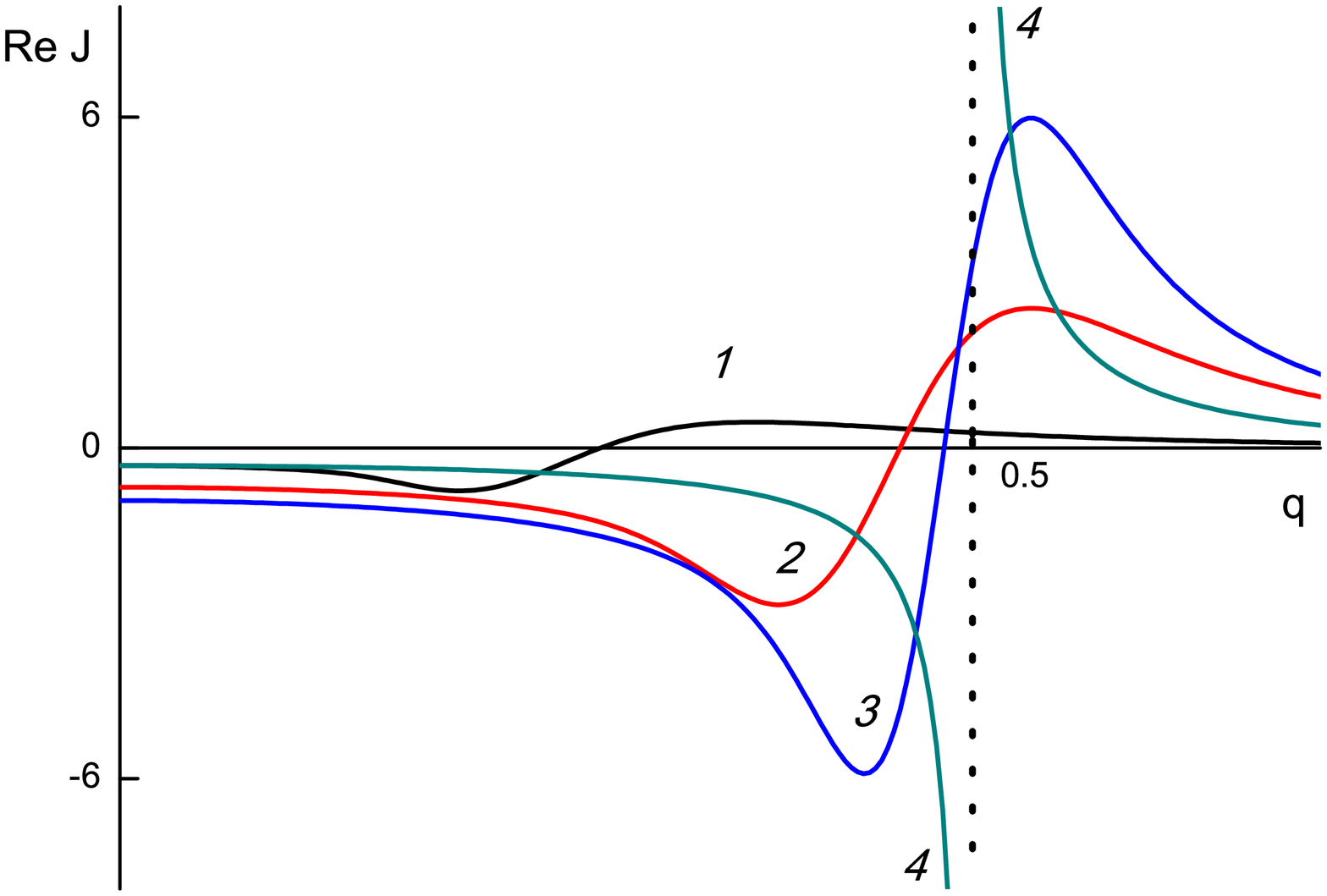}
{\bf Fig. 1.} Real part of density of dimensionless current
in classical Fermi---Dirac plasma (curves 1,2,3) and degenerate Fermi
plasma  (curve 4),
$\Omega=0.5$; curves 1, 2, 3 corres\-pond to values of
dimensionless chemical potential $\alpha=1,5,9$.
\end{figure}

\clearpage

\begin{figure}[t]\flushleft
\includegraphics[width=16.0cm, height=17cm]{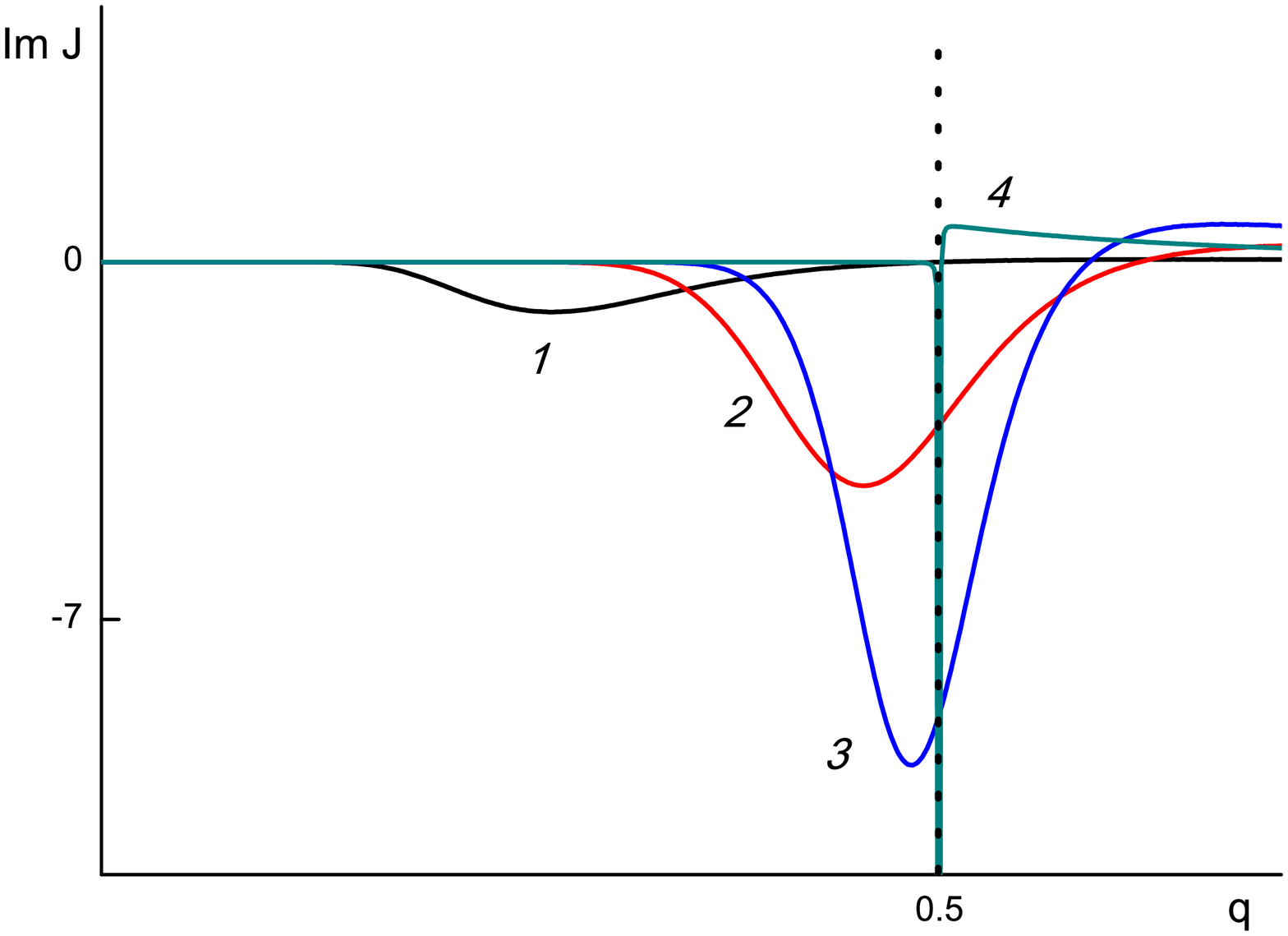}
{\bf Fig. 2.} Imaginary part of density of dimensionless current
in classical Fermi---Dirac plasma (curves 1,2,3) and degenerate Fermi
plasma  (curve 4),
$\Omega=0.5$; curves 1, 2, 3 correspond to values of
dimensionless chemical potential $\alpha=1,5,9$.
\end{figure}

\begin{figure}[t]\flushleft
\includegraphics[width=18.0cm, height=17cm]{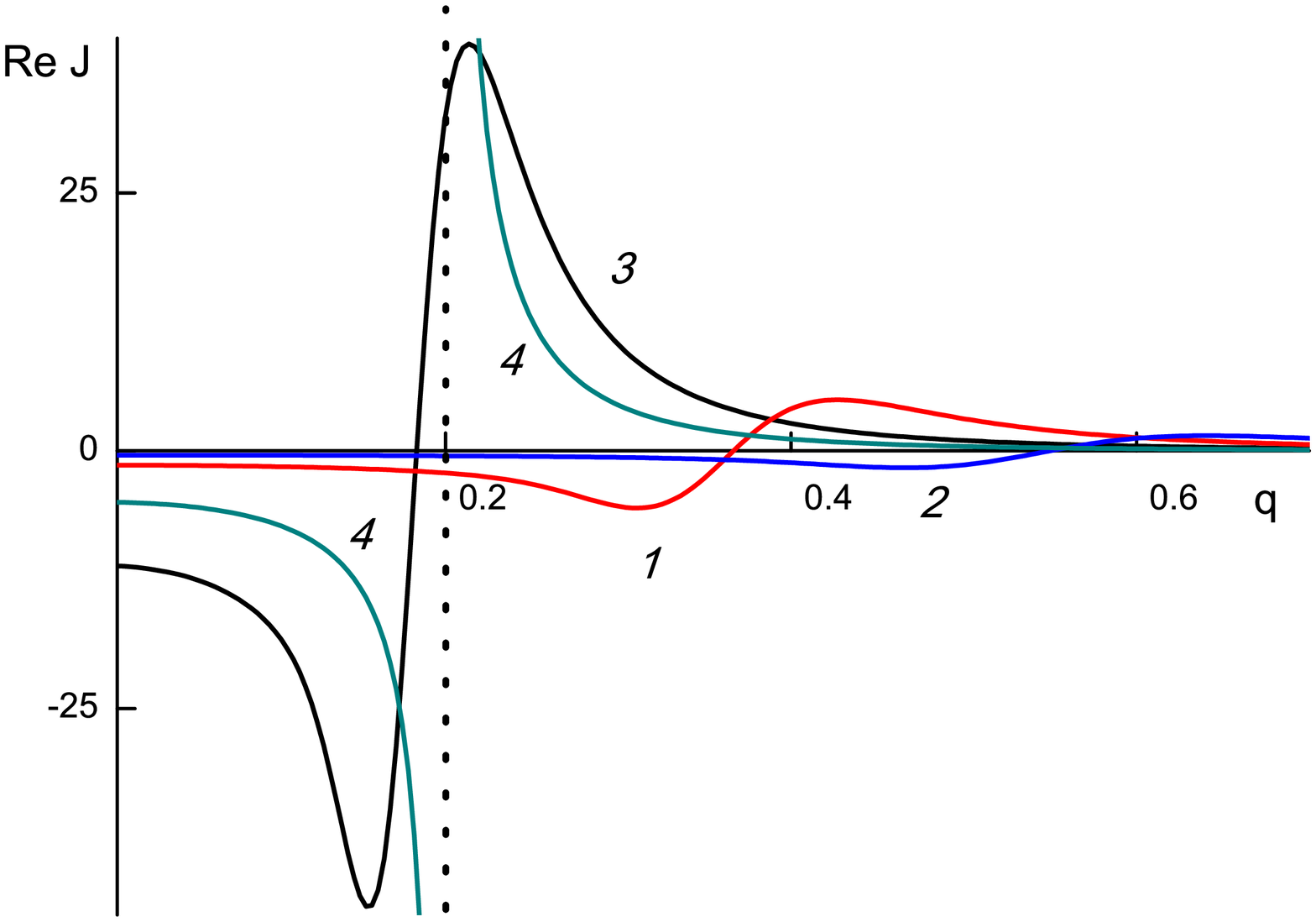}
{\bf Fig. 3.}  Real part of density of dimensionless current
in classical Fermi---Dirac plasma (curves 1,2,3) and degenerate Fermi
plasma  (curve 4) at $\Omega=0.2$, $\alpha=5$
$\Omega=0.5$; curves 1, 2, 3 correspond to values of
dimensionless oscillation frequency $\Omega=0.2,0.4,0.6$.
\end{figure}

\begin{figure}[t]\flushleft
\includegraphics[width=18.0cm, height=16cm]{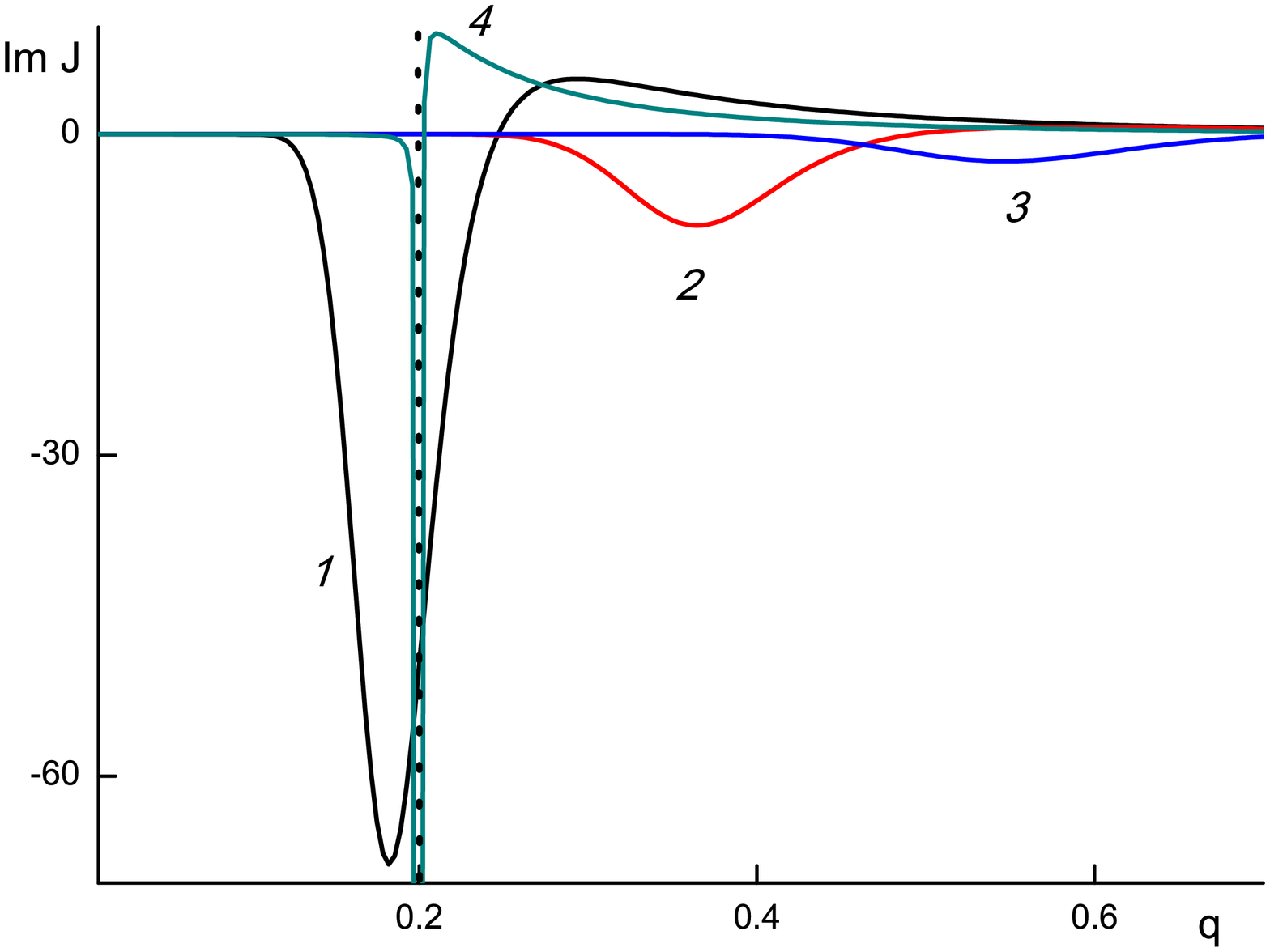}
{\bf Fig. 4.}  Imaginary part of density of dimensionless current
in classical Fermi---Dirac plasma (curves 1,2,3) and degenerate Fermi
plasma  (curve 4) at $\Omega=0.2$, $\alpha=5$
$\Omega=0.5$; curves 1, 2, 3 correspond to values of
dimensionless oscillation frequency $\Omega=0.2,0.4,0.6$.
\end{figure}
\clearpage

\begin{center}
  \bf 2. Degenerate quantum plasma
\end{center}

In our previous work \cite{Lat7}
the following expression for the longitudinal current has been received
$$
\mathbf{j}^{\rm quadr}=\dfrac{2ep_T^3}{(2\pi\hbar)^3}\int \Bigg[
\dfrac{e^2v_T^3\mathbf{P(PA)}^2}{2c^2\hbar^2(\omega-v_T\mathbf{kP})}
\Big(\dfrac{f_0(\mathbf{P+q})-f_0(P)}{\omega-v_T\mathbf{k(P+q}/2)}+$$$$+
\dfrac{f_0(\mathbf{P-q})-f_0(P)}{\omega-v_T\mathbf{k(P-q}/2)}\Big)-
\dfrac{e^2v_T\mathbf{PA}^2}{4mc^2\hbar}\dfrac{f_0(\mathbf{P+q})-
f_0(\mathbf{P-q})}{\omega-v_T\mathbf{kP}}\Bigg]d^3P.
\eqno{(2.1)}
$$

Let us consider the case of degenerate plasmas.

In the formula (2.1) we will carry out replacement of variables of integration
$$
\mathbf{P}\to \dfrac{v_0}{v_T}\mathbf{P},
$$
where $v_0$ is the electron velocity on Fermi--surface.

We receive the following expression for quantity of electric current in
quantum plasma
$$
\mathbf{j}^{\rm quadr}=\dfrac{2ep_0^3}{(2\pi\hbar)^3}\int \Bigg[
\dfrac{e^2v_0^3\mathbf{P(PA)}^2}{2c^2\hbar^2(\omega-v_0\mathbf{kP})}
\Big(\dfrac{f_0(\mathbf{P+q})-f_0(P)}{\omega-v_0\mathbf{k(P+q}/2)}-$$$$-
\dfrac{f_0(P)-f_0(\mathbf{P-q})}{\omega-v_0\mathbf{k(P-q}/2)}\Big)-
\dfrac{e^2v_0\mathbf{PA}^2}{4mc^2\hbar}\dfrac{f_0(\mathbf{P+q})-
f_0(\mathbf{P-q})}{\omega-v_0\mathbf{kP}}\Bigg]d^3P.
\eqno{(2.2)}
$$

As well as earlier, vector equality (2.2) contains one nonzero
component
$$
{j_x}^{\rm quant}=
\dfrac{e^3p_0^3A_y^2}{(2\pi\hbar)^3c^2m^2v_0q^2}\int \Bigg[
\Big(\dfrac{f_0({\bf P+q})-f_0(P)}{P_x+q/2-\Omega/q}-
$$$$
-\dfrac{f_0(P)-f_0({\bf P-q})}{P_x-q/2-\Omega/q}\Big)
\dfrac{P_xP_y^2}{P_x-\Omega/q}+
$$
$$
+\dfrac{q}{2}\dfrac{f_0({\bf P+q})-f_0({\bf P-q})}{P_x-\Omega/q}P_x
\Bigg]d^3P.
\eqno{(2.3)}
$$

Here
$$
f_0({\bf P\pm q})=\Big[1+\exp\Big((P_x\pm q)^2+
P_y^2+P_z^2-\alpha\Big)\Big]^{-1}.
$$

In formula (2.2) $p_0=mv_0$ is the electron momentum on Fermi--surface,
$$
f_0=\dfrac{1}{1+\exp\dfrac{\E_0P^2-\mu}{k_BT}}=
\dfrac{1}{1+\exp\Big(\dfrac{\E_0}{\E_T}P^2-\dfrac{\mu}{\E_T}\Big)},
$$
where $\E_0$ is the electron energy on Fermi--surface, $k_0$ is
the wave Fermi number, $\E_T=\dfrac{mv_T^2}{2}$,
$$
q=\dfrac{k}{k_0},\qquad \Omega=\dfrac{\omega}{k_0v_0}, \qquad
k_0=\dfrac{mv_0}{\hbar}.
$$

Let us notice, that in the limit of zero absolute temperature it is had
$$
 \lim\limits_{T\to 0}\mu =\E_0, \qquad \E_0=\dfrac{mv_0^2}{2},
$$

Hence, in the limit of zero temperature absolute
Fermi---Dirac distribution function passes in the absolute
Fermi's distribution function for degenerate plasmas
$$
\lim\limits_{T\to 0}f_0=\lim\limits_{T\to 0}\dfrac{1}{1+\exp\Big(
\dfrac{\E_0P^2-\mu}{\E_T}\Big)}=
$$
$$
=\Theta\Big(\dfrac{\E_0}{\E_T}(1-P^2)\Big)=
\Theta(1-P^2).
$$

Thus, in the limit of zero temperature the formula (2.3)
will be transformed to the form
$$
{j_x}^{\rm quant}=
\dfrac{e^3p_0^3A_y^2}{(2\pi\hbar)^3c^2m^2v_0q^2}\int \Bigg[
\Big[\dfrac{\Theta[1-(P_x+q)^2-P_y^2-P_z^2]-\Theta[1-P^2]}{P_x+q/2-\Omega/q}-
$$$$
-\dfrac{\Theta[1-P^2]-\Theta[1-(P_x-q)^2-P_y^2-P_z^2]}{P_x-q/2-\Omega/q}\Big]
\dfrac{P_xP_y^2}{P_x-\Omega/q}+
$$
$$
+\dfrac{q}{2}\dfrac{\Theta[1-(P_x+q)^2-P_y^2-P_z^2]-
\Theta[1-(P_x-q)^2-P_y^2-P_z^2)}{P_x-\Omega/q}P_x\Bigg]d^3P.
$$
or, having entered designations
$$
\Theta(P_x\pm q)=\Theta[1-(P_x+q)^2-P_y^2-P_z^2],\qquad
\Theta(P)=\Theta(1-P^2),
$$
we rewrite last equality in the form
$$
{j_x}^{\rm quant}=
\dfrac{e^3p_0^3A_y^2}{(2\pi\hbar)^3c^2m^2v_0}\int \Bigg[
\Big[\dfrac{\Theta(P_x+q)-\Theta(P)}{qP_x+q^2/2-\Omega}+
$$$$
+\dfrac{\Theta(P_x-q)-\Theta(P)}{qP_x-q^2/2-\Omega}\Big]
\dfrac{P_xP_y^2}{qP_x-\Omega}+
\dfrac{\Theta(P_x+q)-
\Theta(P_x-q)}{2(qP_x-\Omega)}P_x\Bigg]d^3P.
\eqno{(2.4)}
$$

The expression facing in integral in (2.4), we will transform with
the help of expression for numerical density and with use
communications of potential and intensity of the electromagnetic field
$$
\dfrac{e^3p_0^3A_y^2}{(2\pi\hbar)^3c^2m^2v_0}=-
\dfrac{3e\omega_p^2 E_y^2}{32\pi^2 p_0\omega^2}=-
\dfrac{e\Omega_p^2}{k_0p_0}E_y^2k\dfrac{3}{32\pi^2\Omega^2q}=
$$
$$
=-\sigma_{\rm l,tr}kE_y^2\dfrac{3}{32\pi^2\Omega^2q}.
$$

Let us designate
$$
I=I(\Omega,q)=\int \Bigg[
\Big[\dfrac{\Theta(P_x+q)-\Theta(P)}{qP_x+q^2/2-\Omega}+
\dfrac{\Theta(P_x-q)-\Theta(P)}{qP_x-q^2/2-\Omega}\Big]
\dfrac{P_xP_y^2}{qP_x-\Omega}+$$$$+
\dfrac{\Theta(P_x+q)-
\Theta(P_x-q)}{2(qP_x-\Omega)}P_x\Bigg]d^3P.
\eqno{(2.5)}
$$

By means of a designation (2.5) we will copy equality (2.4) in the form
$$
j_x^{\rm quant}=-\dfrac{3}{32\pi^2 \Omega^2q}I(\Omega,q)
\sigma_{\rm l,tr}k E_y^2,
$$
or
$$
j_x^{\rm quant}=J_q(\Omega,q)\sigma_{\rm l,tr}k E_y^2,
\eqno{(2.4')}
$$
where
$$
J_q(\Omega,q)=-\dfrac{3}{32\pi^2 \Omega^2q}I(\Omega,q),
$$
and the quantity $I (\Omega, q) $ is defined by equality (2.5).

Let us transform integrals from (2.5), using the obvious linear
replacements of variables
$$
I_1=\int \dfrac{\Theta(P_x+q)P_xP_y^2d^3P}{(qP_x+q^2/2-\Omega)(qP_x-\Omega)}=
\int\dfrac{\Theta(P)(P_x-q)P_y^2d^3P}{(qP_x-q^2/2-\Omega)(qP_x-q^2-\Omega)},
$$
$$
I_3=\int \dfrac{\Theta(P_x-q)P_xP_y^2d^3P}{(qP_x-q^2/2-\Omega)(qP_x-\Omega)}=
\int\dfrac{\Theta(P)(P_x+q)P_y^2d^3P}{(qP_x+q^2/2-\Omega)(qP_x+q^2-\Omega)}.
$$
Let us summarize the first and third integrals
$$
I_1+I_3=\int\dfrac{2P_x(qP_x-\Omega)^2-3q^3(qP_x-\Omega)+P_xq^4}
{[(qP_x-\Omega)^2-q^4/4][(qP_x-\Omega)^2-q^4]}\Theta(P)P_y^3d^3P.
$$
Let us summarize the second and fourth integrals
$$
I_2+I_4=-\int\Bigg[\dfrac{1}{qP_x+q^2/2-\Omega}+
\dfrac{1}{qP_x-q^2/2-\Omega}\Bigg]\dfrac{\Theta(P)P_xP_y^2d^3P}{qP_x-\Omega}=
$$
$$
=-\int\dfrac{2\Theta(P)P_xP_y^2d^3P}{[(qP_x-\Omega)^2-q^4]}.
$$
The sum of first four integrals is equal
$$
I_1+I_2+I_3+I_4=3q^2\Omega\int\dfrac{\Theta(P)P_y^3d^3P}
{[(qP_x-\Omega)^2-q^4/4][(qP_x-\Omega)^2-q^4]}.
$$

In the same way the fifth and sixth integrals give
$$
I_5+I_6=\int\dfrac{\Theta(P_x+q)-
\Theta(P_x-q)}{2(qP_x-\Omega)}P_xd^3P=q\Omega\int\dfrac{\Theta(P)d^3P}
{(qP_x-\Omega)^2-q^4}.
$$

Thus, the integral $I $ is equal
$$
I=3q^2\Omega\int\dfrac{\Theta(P)P_y^3d^3P}
{[(qP_x-\Omega)^2-q^4/4][(qP_x-\Omega)^2-q^4]}+q\Omega\int\dfrac{\Theta(P)d^3P}
{(qP_x-\Omega)^2-q^4}.
$$

Internal integrals in  plane $ (P_y, P_z) $ are equal
$$
\iint\limits_{P_y^2+P_z^2<1-P_x^2}P_y^2dP_ydP_z=\dfrac{\pi}{4}(1-P_x^2)^2,
$$
$$
\iint\limits_{P_y^2+P_z^2<1-P_x^2}dP_ydP_z=\pi (1-P_x^2).
$$

Hence, the integral $I $ is reduced to one-dimensional integral
$$
I=\dfrac{3\pi}{4}q^3\Omega \int\limits_{-1}^{1}
\dfrac{(1-\tau^2)^2d\tau}{[(q\tau-\Omega)^2-q^4/4][(q\tau-\Omega)^2-q^4]}+
\pi q\Omega\int\limits_{-1}^{1}
\dfrac{(1-\tau^2)d\tau}{(q\tau-\Omega)^2-q^4}=
$$
$$
=\dfrac{\pi q\Omega}{4}\int \dfrac{(1-\tau^2)\{q^2[3(1-\tau^2)-q^2]+
4(q\tau-\Omega)^2\}}{[(q\tau-\Omega)^2-q^4/4][(q\tau-\Omega)^2-q^4]}d\tau.
$$

Thus, the longitudinal current in quantum plasma is equal
$$
j_x^{\rm quant}=J_q(\Omega,q)\sigma_{\rm l,tr}kE_y^2,
\eqno{(2.6)}
$$
where
$$
J_q(\Omega,q)=-\dfrac{3}{32\pi \Omega}\int\limits_{-1}^{1}
 \dfrac{(1-\tau^2)\{q^2[3(1-\tau^2)-q^2]+
4(q\tau-\Omega)^2\}}{[(q\tau-\Omega)^2-q^4/4][(q\tau-\Omega)^2-q^4]}d\tau.
$$

Let us copy equality (2.6) in the invariant form
$$
{\bf j}^{\rm quant}=J_q(\Omega,q)\sigma_{\rm l,tr}{\bf kE}^2_{\rm tr}
=J_q(\Omega,q)\sigma_{\rm l,tr}\dfrac{\omega}{c}[{\bf E,H}],
$$
where ${\bf E}_{\rm tr}$ is the transversal electrical field,
entered above.

Let us consider the case of small values of the wave vector. We take
equality $ (2.4 ') $. We will notice, that at small values $q $
the square bracket in (2.5) is equal to zero in linear approximation.
Now we use the linear decomposition
$$
\Theta(P_x\pm q)=\Theta(1-P^2)+\delta(1-P^2)(\mp 2P_xq).
$$
Hence, in linear approximation we have
$$
\dfrac{\Theta(P_x+q)-\Theta(P_x-q)}{2(qP_x-\Omega)}=
\dfrac{q}{\Omega}P_x\delta(P-1).
$$

Now it agree $ (2.4 ') $ we receive
$$
j_x^{\rm quant}=-\sigma_{\rm l,tr}kE_y^2\cdot \dfrac{1}{8\pi \Omega^3}.
\eqno{(2.7)}
$$

Let us notice, that expression (2.7) in accuracy coincides
with expression of the longitudinal current (1.15) for
classical degenerate plasma.

So, by us it is shown, that at small values of the wave vector
longitudinal current in classical and quantum degenerate plasma
is calculated under the same formula (1.14).

On Figs. 5 -- 9 the behaviour of real  and
imaginary parts of degenerate classical and quantum
plasma is graphically investigated.
Let us notice, that the real part of a current in quantum plasma has
at first  minimum (at $q <\Omega $), and then has  maximum (at
$q> \Omega $). The imaginary part of the current in quantum plasma
has the minimum at any values $ \Omega $.

On Figs. 5 and 6 we will result comparison accordingly of real
(Fig. 5) and imaginary (Fig. 6) parts of the dimensionless part of
density of the longitudinal current of classical and quantum plasma.

From Figs. 5 and 6 it is visible, that at small values dimensionless
wave number $q $ quantity of the real part of the longitudinal
current of classical and quantum plasma the close friend to the friend and
coincide in a limit at $k\to 0$. This fact is a consequence
coincidence of equalities (1.14) and (2.7).
At $q\to \infty $ values and
real, and imaginary parts of the longitudinal current in quantum and
classical plasma approach.

On Figs. 7 and 8 we will represent behaviour of real (Fig. 7) and
imaginary (Fig. 8) parts of the longitudinal current in quantum plasma in
dependences on dimensionless wave number $q $ at the various
values of dimensionless frequency of an electromagnetic field $ \Omega $.

At decrease  of quantity $ \Omega $
growth of values of quantity of real and imaginary
parts of the longitudinal current is observed.

On Fig. 9 the imaginary part of density of the dimensionless
current is represented in the case $ \Omega=0.2$;
curves 1 and 2 answer accordingly to degenerate classical and quantum plasma.

\clearpage

\begin{figure}[t]\center
\includegraphics[width=16.0cm, height=10cm]{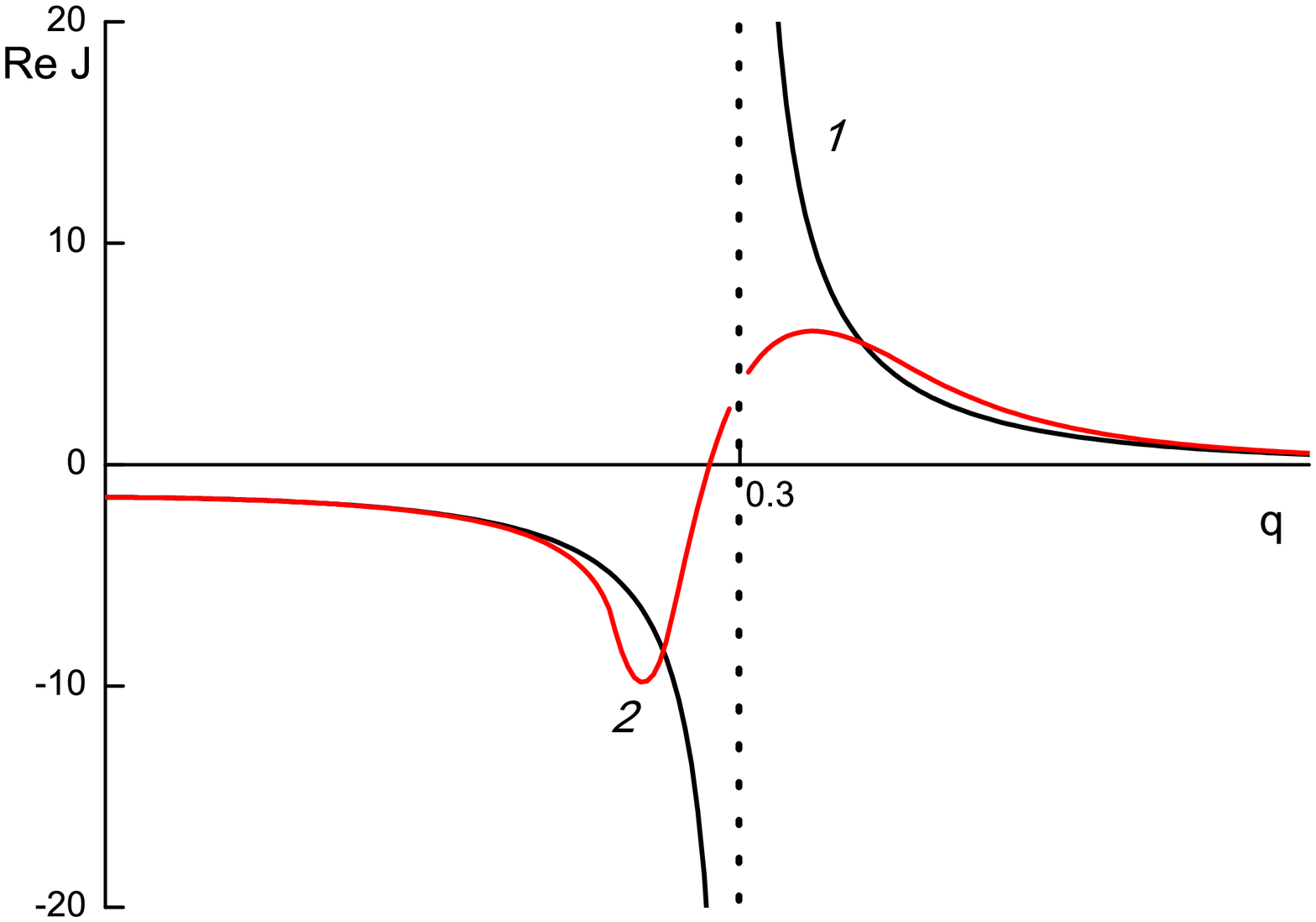}
{\bf Fig. 5.}  Real part of density of dimensionless current,
$\Omega=0.3$; curves 1 and 2 correspond to classical and quantum plasma.
\includegraphics[width=16.0cm, height=10cm]{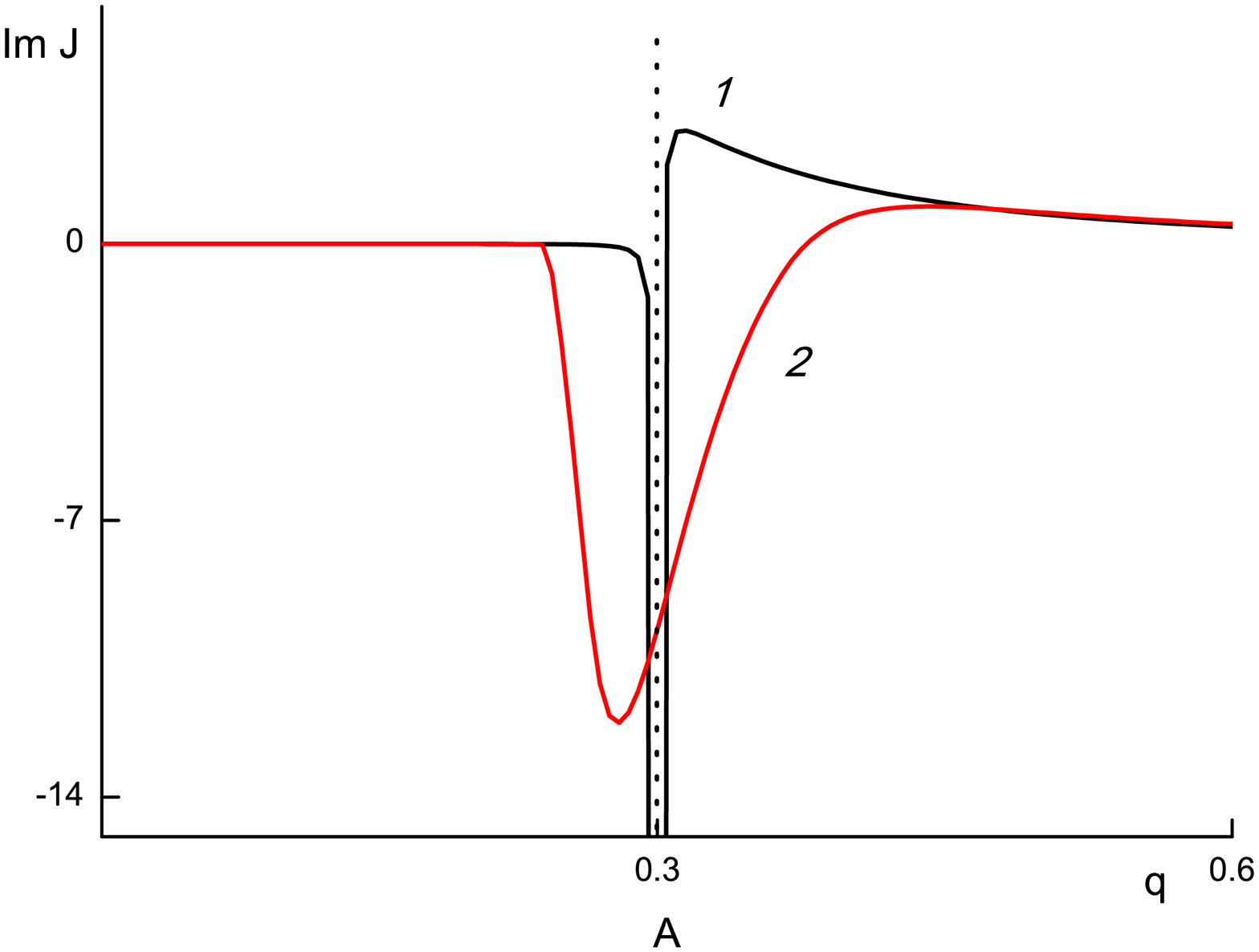}
{\bf Fig. 6.}  Imaginary part of density of dimensionless current,
$\Omega=0.5$; curves 1 and 2 correspond to classical and quantum plasma.
\end{figure}

\begin{figure}[t]\center
\includegraphics[width=16.0cm, height=10cm]{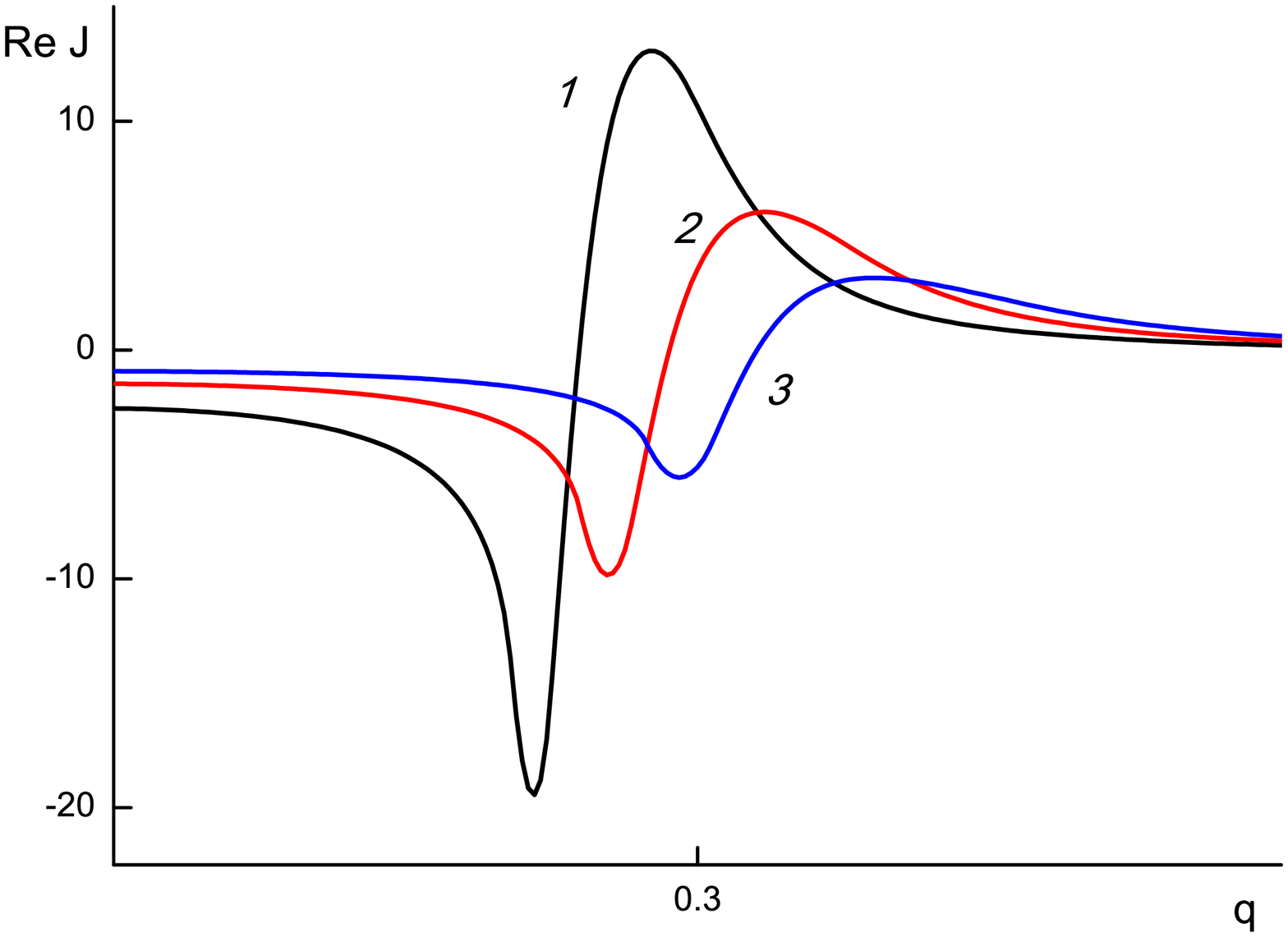}
{\bf Fig. 7.}  Real part of density of dimensionless current of quantum
plasma; curves 1, 2 and 3 correspond to values of dimensionless
ocscillation frequency of electromagnetic field
$\Omega=0.25, 0.3, 0.35$.
\includegraphics[width=16.0cm, height=10cm]{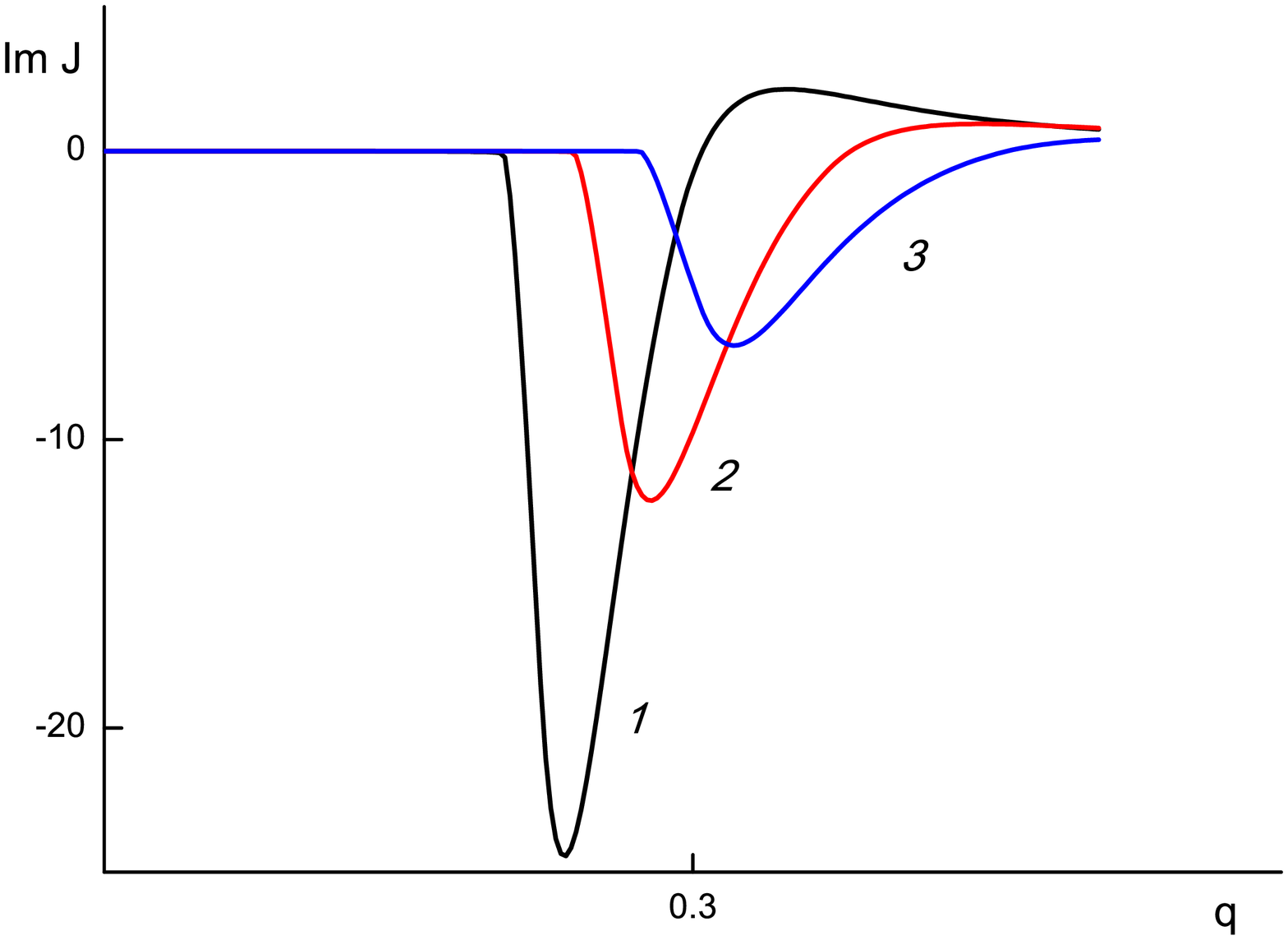}
{\bf Fig. 8.}  Imaginary part of density of dimensionless current of quantum
plasma; curves 1, 2 and 3 correspond to values of dimensionless
ocscillation frequency of electromagnetic field
$\Omega=0.25, 0.3, 0.35$.
\end{figure}

\begin{figure}[t]\center
\includegraphics[width=20.0cm, height=15cm]{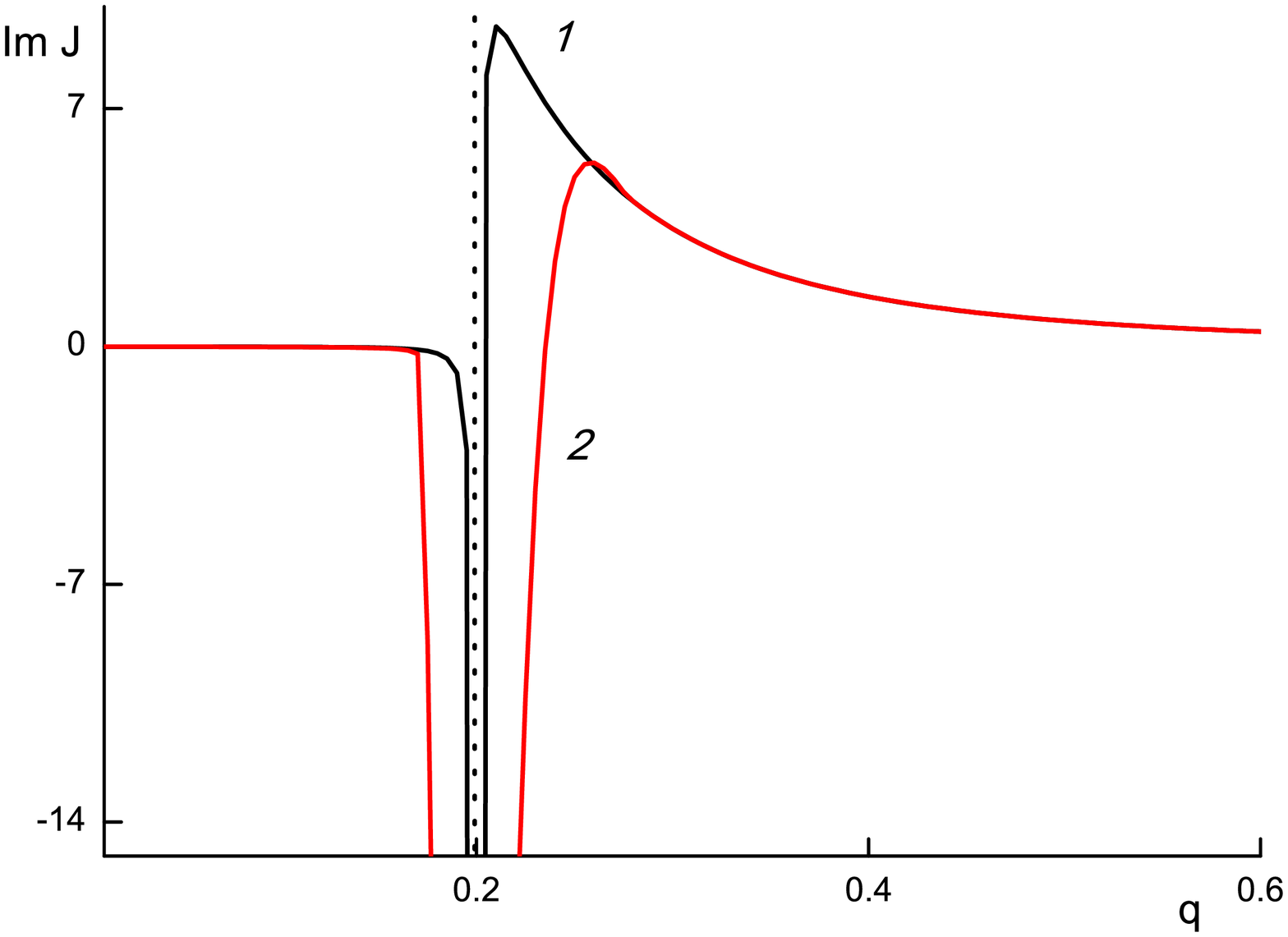}
{\bf Fig. 9.} Imaginary part of density of dimensionless current,
$\Omega=0.2$; curves 1 and 2 correspond to classical and quantum plasma.
\end{figure}

\clearpage

\begin{center}
\bf 3. Conclusion
\end{center}

In the present work the nonlinear analysis of interaction
of electromagnetic wave with classical and quantum degenerate
plasma is carried out.
It is used square-law on intensity quantity of
electric field the decomposition of distribution  function and
Vlasov equation  in case of the classical
degenerate  plasmas.
And also it is used square-law on vector potential  quantity
of electromagnetic field decomposition of quantum distribution
function and Wigner equation in case of the quantum
degenerate plasmas. It is shown, that besides the known transversal
current both in classical and in quantum plasma it is generated
the longitudinal current. Research real and imaginary
parts of the longitudinal current is carried out.

In the following work we will consider generation of
longitudinal current by transversal electromagnetic field
in collisional classical plasma.


\begin{thebibliography}{99}

\bibitem{Gins}
{\it Ginsburg V.L., Gurevich A.V.}
The nonlinear phenomena in the plasma which is
in the variable electromagnetic
field//Uspekhy Fiz. Nauk, {\bf 70}(2) 1960; p. 201-246 (in
Russian).

\bibitem{Zyt2}{\it Kovrizhkhykh L.M. and Tsytovich V.N.}
Effects of transverse electromagnetic wave decay in a
plasma//Soviet physics JETP. 1965. V. 20. \No 4, 978-983.

\bibitem{Zyt}{\it Zytovich V.N.} Nonlinear effects in plasmas//
Uspekhy Fiz. Nauk, {\bf 90}(3) 1966; p. 435-489 (in Russian).

\bibitem{Zyt3} {\it Zytovich V.N.} Nonlinear effects in plasmas.
Moscow. Publ. Leland. 2014. 287 p. (in Russian).

\bibitem{Shukla1}{\it Shukla P. K. and Eliasson B.}
Nonlinear aspects of quantum
plasma physics //
Uspekhy Fiz. Nauk, {\bf 53}(1) 2010;
[V. 180. No. 1, 55-82 (2010) (in Russian)].

\bibitem{Shukla2} {\it Eliasson B. and Shukla P. K.}
Dispersion properties of
electrostatic oscillations in quantum plasmas //
arXiv:0911.4594v1 [physics.plasm-ph] 24 Nov 2009, 9 pp.

\bibitem{Atwal}{\it  Atwal G. S. and  Ashcroft N. W.}
{Relaxation of an Electron System : Conserving Approximation}//
arXiv:cond-mat/0107348v4 [cond-mat.str-el] 18 Sep 2002.

\bibitem{Lat1}{\it Latyshev A.V. and Yushkanov A.A.}
Transverse Electric Conductivity in Collisional Quantum Plasma//
Plas\-ma Physics Report, 2012, Vol. 38, No. 11, pp. 899--908.

\bibitem{Mermin} {\it Mermin N. D.}
{ Lindhard Dielectric Functions in the Relaxation--Time Approximation}.
Phys. Rev. B. 1970. V. 1, No. 5. P. 2362--2363.

\bibitem{Lat2}{\it Latyshev A. V. and Yushkanov A. A.}
Transverse electrical conductivity of a quantum collisional
plasma in the Mermin approach // Theor. and Math. Phys., {\bf
175}(1): 559--569 (2013).

\bibitem{Lat3}{\it Latyshev A. V. and Yushkanov A. A.}
Longitudinal Dielectric Permeability of a
Quntum Degenerate Plasma with a Constant Collision Frequency//
High Temperature, 2014, Vol. 52, \No 1, pp.
128--128.

\bibitem{Lat4}{\it Latyshev A. V. and Yushkanov A. A.}
Longitudinal electric conductivity in a quantum
plasma with a variable collision frequency in the framework of
the Mermin approach// Theor. and Mathem. Physics, {\bf 178}(1):
131-142 (2014).

\bibitem{Lat5}{\it Latyshev A. V. and Yushkanov A. A.}
Transverse Permittivity of Quantum Collisional
Plasma with an Arbitrary Collision Frequency//ISSN 1063-780X,
Plasma Physics Reports, 2014, Vol. 40, No. 7, pp. 564-571.


\bibitem{Lat6}{\it Latyshev A. V. and Yushkanov A. A.}
Nonlinear phenomena of generation of longitudinal electric
current by transversal electromagnetic field in plasmas//
arXiv:1502.04581v1 [phy\-si\-cs.plasm-ph], 16 Feb 2015, 16 p.


\bibitem{Lat7}{\it Latyshev A. V. and Yushkanov A. A.}
Generation of the longitudinal current by the
transversal electromagnetic field in classical and quantum
plasmas//arXiv: 1503.02102 [physics.plasm-ph] 6 Mar 2015, 27 p.

\bibitem{Lat8}{\it Latyshev A. V. and Yushkanov A. A.}
Generation of longitudinal
electric current by transversal electromagnetic field in
Maxwellian plasmas// arXiv: 1503.04478 [physics.plasm-ph] 15 Mar 2015, 18  p.


\end{thebibliography}
\end{document}